\documentclass[pre,superscriptaddress,showkeys,showpacs,twocolumn]{revtex4-1}
\usepackage{graphicx}
\usepackage{latexsym}
\usepackage{amsmath}
\begin {document}
\title {Robust criticality of Ising model on rewired directed networks}
\author{Adam Lipowski}
\affiliation{Faculty of Physics, Adam Mickiewicz University, Pozna\'{n}, Poland}
\author{Krzysztof Gontarek}
\affiliation{Faculty of Physics, Adam Mickiewicz University, Pozna\'{n}, Poland}
\author{Dorota Lipowska}
\affiliation{Faculty of Modern Languages and Literature, Adam Mickiewicz University, Pozna\'{n}, Poland}
%%%%%%%%%%%%%%%%%%%%%%%%%%%%%%%%%%%%%%%%%%%%%%%%%%%%%%%%%%%%%%%%%%%%%%%%%%%%%
\begin {abstract} We show that preferential rewiring, which is supposed to mimick the behaviour of financial agents, changes a directed-network Ising ferromagnet with a single critical point into a model with robust critical behaviour. For the non-rewired random graph version, due to a constant number of out-links  for each site, we write a simple mean-field-like equation describing the behaviour of magnetization; we argue that it is exact and support the claim  with extensive Monte Carlo simulations. For the rewired version, this equation is obeyed only at low temperatures. At higher temperatures, rewiring leads to strong heterogeneities, which apparently invalidates mean-field arguments and induces large fluctuations and divergent susceptibility. Such behaviour is traced back to the formation of a relatively small core of agents which influence the entire system.
\end{abstract}
\pacs{} \keywords{}

\maketitle
\section{Introduction}
To understand a complex behaviour  of financial markets is one of the main objectives of econophysics. Fat-tail non-Gaussian fluctuations, volatility clustering or rapid decay of autocorrelations of returns characterize most of the financial markets, suggesting that these stylized facts \cite{stylized} have some  more fundamental explanation. Searching for such an explanation, one can  resort to the approach particularly suited for physicists, namely agent modeling \cite{agent}. In the spirit of statistical mechanics, one considers a collection of agents involved in interactions resembling functioning of financial markets. Since buying and selling are the most important activities of such agents, a number of models of financial markets bear some similarity to the two-state percolation \cite{cont} or Ising-like models \cite{ising}. 

An important agent's interaction is mimicking some other agents' behaviour, which suggests a similarity to ferromagnetic systems. However, ferromagnets typically exhibit rather small fluctuations, which is much different from the behaviour of finacial markets. Ferromagnets exhibit large fluctuations only at the critical point separating ferromagnetic and paramagnetic phases. To place the system at the critical point requires, however, a fine tuning of control parameters. On the other hand, financial markets seem to be more robust with strong fluctuations appearing without any tuning of parameters. Some models were proposed, where mimicking other agents' behaviour is compensated with the tendency to be in the minority \cite{bornholdt} or where agents with more complex strategies were used \cite{lux}. They do reproduce some of the stylized facts but their considerable complexity hinders deeper understanding.

Apparently, the analogy with simple ferromagnetic systems is not sufficient to  model financial markets and one should search for more suitable extensions. 
In our opinion, an important ingredient of models of financial markets should be the possibility to choose and sometimes also change neighbours that a given agent would like to mimick.
The objective of the present paper is to implement such a rewiring mechanism and to show that it drastically affects the behaviour of the model.  When the neighbours to be mimicked are selected at random and kept fixed, the model behaves as an ordinary ferromagnet with ferromagnetic and paramagnetic phases separated at a critical  point. However, when agents might switch the neighbours and  preferentially select those they consider as more influential, the system generically exhibits divergent fluctuations. Such behaviour indicates that preferential rewiring induces a robust criticality, which  is a required feature of stock-market models \cite{tanaka}. We also examine the mechanism leading to the robust criticality.

\section{Model without rewiring}
In our model we consider $N$ agents represented by spin-like variables $s_i=\pm 1, \ i=1,2\ldots, N$. 
At each time step $t$, each agents decides whether to buy ($s_i=1$) or sell ($s_i=-1$) an asset. To make the decision, an agent tries to mimick the behaviour of its neighbours and the model evolves according to the heat-bath dynamics:
\begin{equation}
s_i(t+1)=\left\{ \begin{array}{ll} 1 & {\rm \ with \ probability\ } p=\frac{1}{1+\exp[-2h_i(t)/T]}\\
 -1 & {\rm \ with \ probability\ } 1-p
 \end{array}\right.
\label{heat-bath}
\end{equation}
where 
\begin{equation}
h_i(t)=\sum_{j} s_j(t)
\label{field}
\end{equation}
is the local field acting on a given agent $i$ and summation in Eq.~(\ref{field}) is over its neighbours. The control parameter~$T$ is the analogue of the temperature in the magnetic Ising model and determines the level of fluctuations in the decision process. 

The neighbourhood of a given agent is set randomly, namely, each agent has a fixed number of $z$ randomly selected neighbours, which it interacts with via the local field. The neighbouring relation is not necessarily symmetric: if agent~$j$ enters the expression for the local field  of agent~$i$, it does not imply that agent~$i$ enters the expression for the local field  of agent~$j$. In other words, agents are nodes of a directed random network and each node has $z$ out-links (arrows point at the nodes that contribute to the local field). The number of in-links of a given agent, which specifies how many agents it influences, is not fixed and it can vary among agents (of course, the average over all agents equals $z$). Equal numbers of out-links and unequal numbers of in-links constitute an important feature of our model, which we will refer to as the out-homogeneity. 

Taking into account the spin variables, the above rules define actually an Ising ferromagnet on a directed random graph. Models of this kind were already analysed and shown to exibit an ordinary ferro-para phase transition belonging to the mean-field universality class \cite{directed-ising}. 

In the following, we present a more detailed analysis of our model for $z=4$. Due to the out-homogeneity, one can write a relatively simple equation, which governs the evolution of magnetization. Let $P_{i}(t)$ denote the probability that agent $i$ at time $t$ takes the value $s_i=1$. Assuming that $P_{i}(t)$ is spatially homogenous and does not depend on~$i$, from the heat-bath rules we obtain that
\begin{equation}
P(t+1)=\sum_{k=0}^{4} {4 \choose k} P^k(t)[1-P(t)]^{4-k}\frac{1}{1+\exp{[-4(k-2)/T]}} .
\label{mfa}
\end{equation}
Of course, Eq.~(\ref{mfa}) can be easily rewritten in terms of magnetization ($m(t)=2P(t)-1$), which is  common in Ising-model studies.
In the steady-state limit ($t\rightarrow \infty$).  Eq.~(\ref{mfa}) becomes a 4-th order polynomial equation, which can be easily solved numerically (and with some more effort even analytically). Moreover, the critical temperature $T_c$ can be found using the standard procedure of expanding the $t\rightarrow \infty$ limit of Eq.~(\ref{mfa}) in the vicinity of the critical point. Elementary calculations reveal that $T_c$ obeys 
\begin{equation}
2=\tanh{(4/T_c)}+2\tanh{(2/T_c)} .
\label{tc}
\end{equation}
The solution of the  equation above can be written as
\begin{equation}
T_c=\frac{4}{\ln{(\frac{1+x}{1-x})}} ,
\label{tc1}
\end{equation}
where
\begin{equation}
x=\frac{1}{3}\Bigg[1-5\sqrt[3]{\frac{2}{11+3\sqrt{69}}}+\sqrt[3]{\frac{1}{2}(11+3\sqrt{69})} \Bigg] .
\label{x}
\end{equation}
We thus obtain $T_c \approx 3.08982$, approximately.

\begin{figure}
\includegraphics[width=3cm]{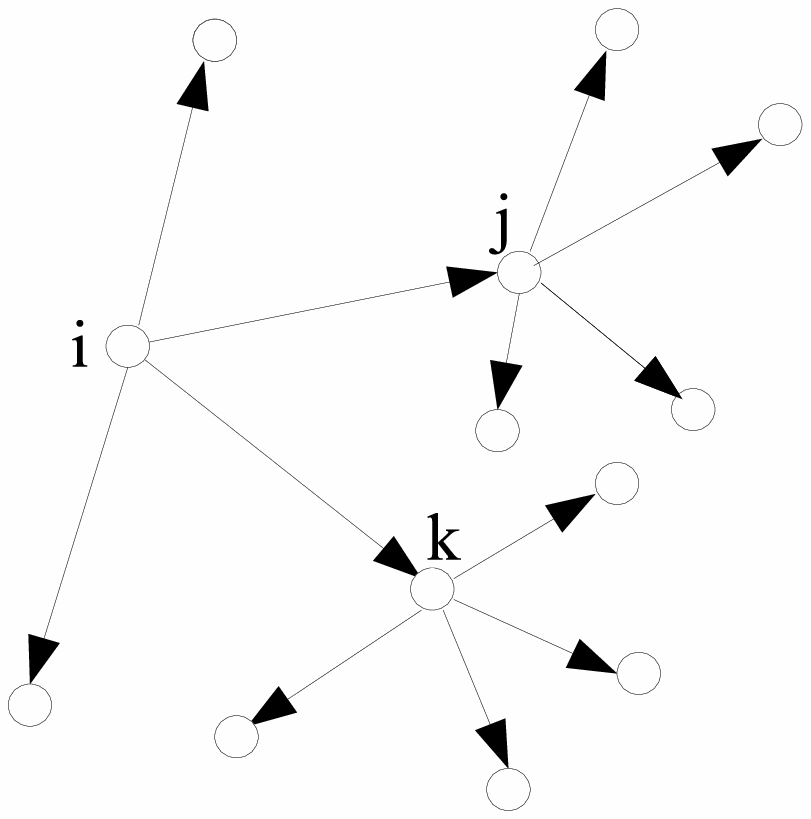}
\caption{In a directed random graph, neighbours $j$ and $k$ of node $i$ are not more correlated than any other two randomly selected nodes.}
\label{ijk}
\end{figure}

The factorized form of the probabilities suggests that Eq.~(\ref{mfa}) is nothing more than the mean-field equation for our model and thus it is only approximate. This would certainly be the case for undirected graphs, where neighbours $j$ and $k$ of agent $i$ are  strongly correlated (since $i$  contributes to the local fields of both $j$ and $k$). 
For undirected random graphs, some insight into the behaviour of the Ising model can obtained using a replica method \cite{replica} or some recurrence relations based on the similarity of random graphs to Cayley trees \cite{dorog}.
On the other hand, in directed networks even though $j$ and $k$ are neighbours of~$i$,  they are not more correlated than any other two randomly selected nodes (Fig.~\ref{ijk}). Since the graph is sparse, we expect that in the limit $N\rightarrow\infty$, such correlations are negligible, and consequently, the factorization in Eq.~(\ref{mfa}) should be legitimate.

%%%%%%%%%%%%%%%%%%%%%%%%%%%%%%%%%%%%%%%%%%%%%%%
%%%%%%%%%%%%%%%%%%%%%%%%%%%%%%%

\begin{figure}
\includegraphics[width=\columnwidth]{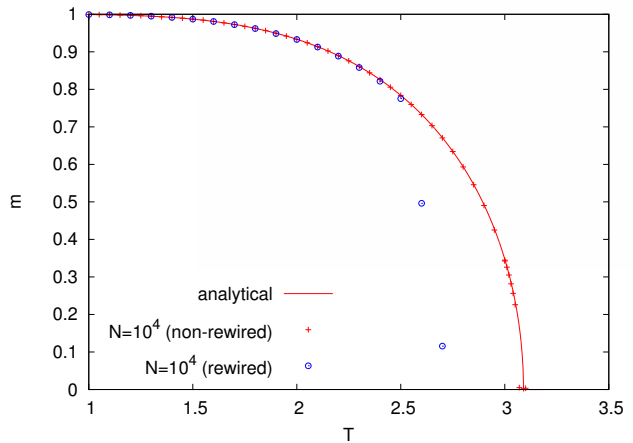}
\caption{Temperature dependence of the magnetization $m$ for the rewired and non-rewired models ($z=4$). The data are obtained from Monte Carlo simulations and are compared with the numerical solution of Eq.~(\ref{mfa}). Simulation and equilibration times were equal to $10^4$ Monte Carlo steps.}
\label{magnet}
\end{figure}

\begin{figure}
\includegraphics[width=\columnwidth]{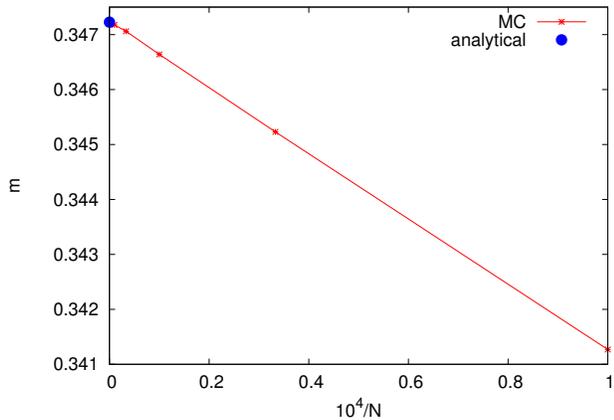}
\caption{Magnetization $m$ as a function of the inverse of size calculated for the non-rewired model of size $N=10^4$, $3\cdot 10^4$, $10^5$, $3\cdot 10^5$, $10^6$, and  $3\cdot 10^6$,  with $z=4$ and $T=3$. Simulation and equilibration times were equal to $10^7$ and $10^4$ Monte Carlo steps, respectively. In the limit $N\rightarrow\infty$, a perfect agreement  with the solution of Eq.~(\ref{mfa}) can be seen.}
\label{check}
\end{figure}

Monte Carlo simulations of our model confirm the above analysis (Fig.~\ref{magnet}). Calculating the magnetization  for $N=10^4$ and $z=4$, we find it in a very good agreement with $m=2P(t=\infty)-1$ obtained from the numerical solution of the steady-state limit of Eq.~(\ref{mfa}).  
%%%%%%%%%%%%%%%%%%%%%%%%%%%%%%%%%%%%%%%%%%%%%%%
%%%%%%%%%%%%%%%%%%%%%%%%%%%%%%%
For $T=3$, we made much more extensive calculations (Fig.~\ref{check}). The linear extrapolation $N\rightarrow\infty$ based on simulations for $N\leq 3\cdot 10^6$ gives $m=0.34723(2)$, which is in perfect agreement with $m=0.347225\ldots$ obtained from the numerical solution of Eq.~(\ref{mfa}). In our opinion, such agreement strongly supports the claim that Eq.~(\ref{mfa}) is exact (at least in the limit $t\rightarrow\infty$).
%%%%%%%%%%%%%%%%%%%%%%%%%%%%%%%%%%%%%%%%%%%%%%%
%%%%%%%%%%%%%%%%%%%%%%%%%%%%%%%

\section{Model with rewiring}
The model analysed in the previous section behaves similarly to some other Ising-like models with ferromagnetic and paramagnetic phases  separated at the critical point.
Our primary motivation is to modify such ordinary ferromagnets so that they would resemble the behaviour of financial markets, at least to some extent.
We are particularly interested in supplanting a fine-tuned critical point with a more generic critical behaviour, which would exist in some, possibly large, temperature range.
So far our agents make the decision to buy or sell based on the observation of their $z$ neighbours, and the assignement of these neighbours is fixed during the entire evolution of the model. In the present section we modify this rule and allow to change the neighbours. The rewiring we use is preferential: each agent has its status equal to the number of in-links that are (currently) attached to it. The selection of a new neighbour takes place with probability proportional to its status \cite{liproulette}. A single step of the dynamics of our model is thus defined as follows:
\begin{itemize}
\item update spin variables $S_i$ ($i=1, 2,\ldots, N$) according to the heat-bath algorithm (\ref{heat-bath}).
\item rewire each agent selecting preferentially anew its $z$ out-links.  
\end{itemize}

Since we keep the dynamics of spin variables basically unchanged, one might expect that Eq.~(\ref{mfa})  still describes the behaviour of our model. Monte Carlo simulations show that to some extent this is indeed the case (Fig.~\ref{magnet}) and a very good agreement with Eq.~(\ref{mfa}) can be seen over much of the temperature range. However, close to the critical point $T=T_c$, the rewired model shows much lower and perhaps zero magnetization. It would be desirable to understand the reasons why Eq.~(\ref{mfa}) is no longer obeyed at higher temperatures. Possible explanations include appearance of correlations (that we argued are negligible in the non-rewired case) or a breakdown of homogeneity (which is also one of the assumptions leading to Eq.~(\ref{mfa})). 
Some efforts to understand the origin of this behaviour will be made in the next section.

\begin{figure}
\includegraphics[width=\columnwidth]{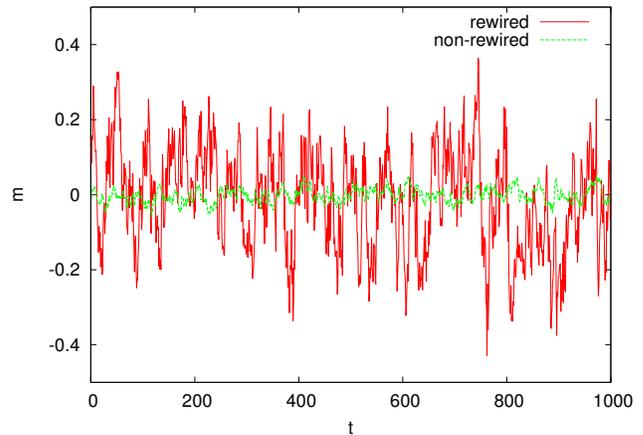}
\caption{Time dependence of magnetization $m$ for the rewired and non-rewired models ($z=4$, $N=10^4$). Simulations were made for $T=4$, which for the non-rewired model is deep in the paramagnetic phase.}
\label{timer0r1}
\end{figure}

\begin{figure}
\includegraphics[width=\columnwidth]{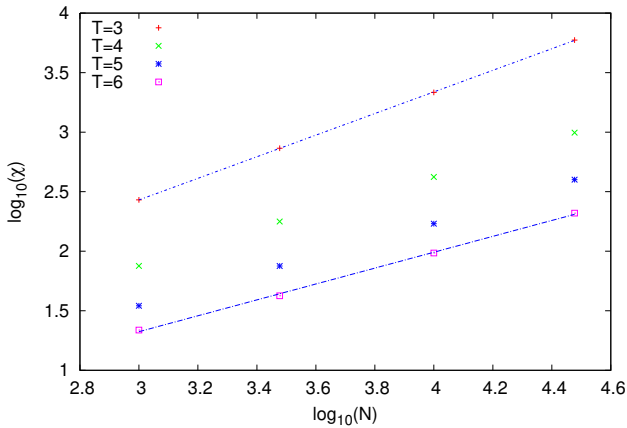}
\caption{Size dependence of susceptibility $\chi$ for the rewired model ($z=4$). The power-law fit $\chi\sim N^{\alpha}$ shows that $\alpha$ varies from 0.67 for $T=6$ up to 0.91 for $T=3$.}
\label{susc}
\end{figure}

What is even more interesting, the magnetization in the rewired version shows large fluctuations also at temperatures much higher than $T_c$ (Fig.~\ref{timer0r1}).
%%%%%%%%%%%%%%%%%%%%%%%%%%%%%%%%%%%%%%%%%%%%%%%
%%%%%%%%%%%%%%%%%%%%%%%%%%%%%%%
To measure these fluctuations more quantitatively, we calculated the susceptibility $\chi$ that up to the temperature factor is equal to the variance of magnetization $\chi = \frac{1}{N}[\langle (\sum_{i=1}^N S_i)^2 \rangle-\langle \sum_{i=1}^N S_i \rangle^2]$. Numerical values indicate that as a function of system size $N$ the susceptibility diverges as $\chi \sim N^\alpha$, where $\alpha\sim 0.67-0.91$ depends slightly on temperature (Fig.~\ref{susc}). Such behaviour is observed in a large temperature range ($3\leq T\leq 6$) for the system size $10^3\leq N\leq 3 \cdot 10^4$ . The divergence of susceptibility indicates that the model exhibits a robust critical behaviour. Together with data from Fig.~\ref{magnet}, this suggests that the model with rewiring has two phases: low-temperature, which is ferromagnetic, and high-temperature, which is critical. It is difficult for us to locate precisely the transition point between these two phases. For longer simulations, it seems to shift slightly toward  lower temperatures. Moreover, one cannot exclude that at sufficiently large temperature the critical phase will be replaced with the paramagnetic one (having much smaller fluctuations).
%%%%%%%%%%%%%%%%%%%%%%%%%%%%%%%%%%%%%%%%%%%%%%%
%%%%%%%%%%%%%%%%%%%%%%%%%%%%%%%

The critical behaviour in our model is also robust with respect to the frequency of rewiring. We made simulations with rewiring taking place,  e.g., with probability 0.1 (i.e., with probability 0.9, the out-links of a given agent at a given step were left  unchanged). Such modification slows down the dynamics but retains the power-law divergence of the susceptibility.
%%%%%%%%%%%%%%%%%%%%%%%%%%%%%%%%%%%%%%%%%%%%%%%%%%%%%%%%%
\section{Dynamics of rewiring}
\begin{figure}
\includegraphics[width=6cm]{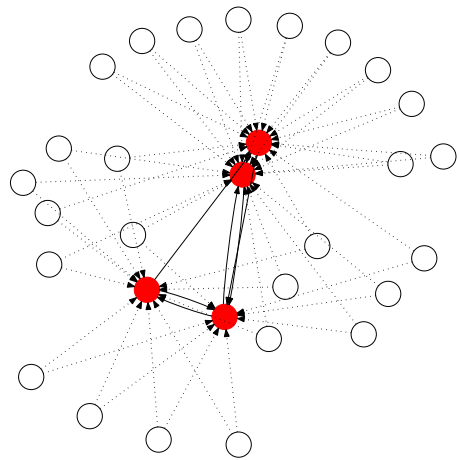}
\caption{Network structure after simulations of $t=10^3$ steps for $N=30$ agents and $z=2$ \cite{graphviz}.  Nodes with no in-links (open circles) represent agents which do not inluence decisions of any other agent (links that go out from them are drawn with dotted lines).  Nodes with some in-links (filled red circles) represent the only  agents that influence other agents (their out-links are drawn with solid lines).}
\label{conf-full}
\end{figure}

\begin{figure}
\includegraphics[width=6cm]{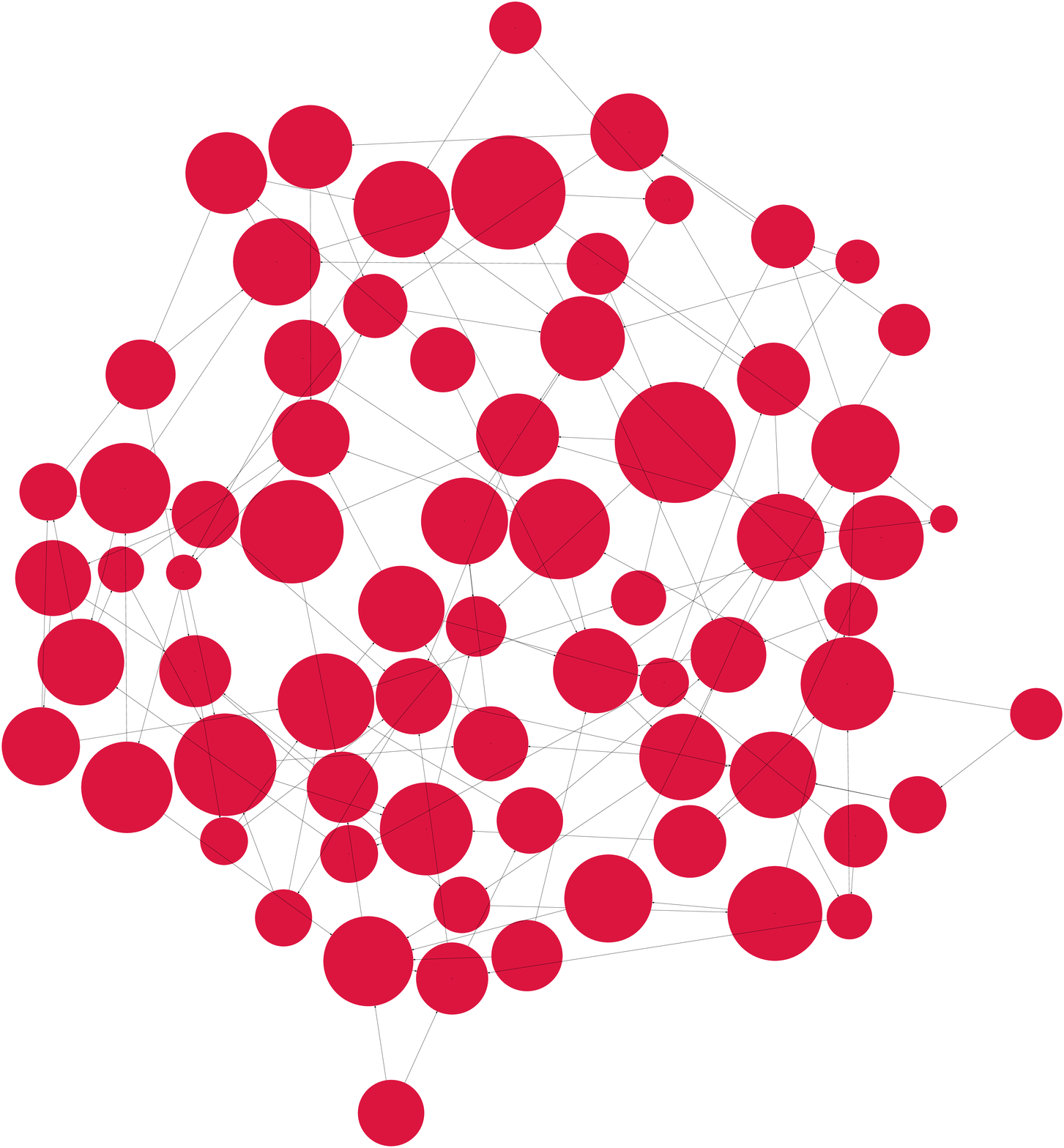}
\caption{Network structure after simulations of $t=10^4$ steps  for $N=10^3$ agents and $z=2$ \cite{graphviz}. Agents with no in-links are omitted. The size of a circle is proportional to the number of in-links.}
\label{conf-core}
\end{figure}

To get some understanding of our model, we looked at the structure of the network that emerges during the rewiring.
Let us notice that rewiring is not affected by spin variables and thus might be considered as an independent process (but not vice versa---spin dynamics depends of course on the structure of the network). Some insight is already obtained from simulations of a small system (Fig.~\ref{conf-full}). One can notice that most agents have no in-links and thus they do not influence any other agent. There is only a small core of agents which are responsible for the decision formation of the other agents. Such structure appears also for larger systems (Fig.~\ref{conf-core}). One can notice a substantial heterogeneity of the resulting core as for the number of agents that a given agent is influencing.
%%%%%%%%%%%%%%%%%%%%%%%%%%%%%%%%%%%%%%%%%%%%%%%
%%%%%%%%%%%%%%%%%%%%%%%%%%%%%%%

%%%%%%%%%%%%%%%%%%%%%%%%%%%%%%%%%%%%%%%%%%%%%%%
%%%%%%%%%%%%%%%%%%%%%%%%%%%%%%%

\begin{figure}
\includegraphics[width=\columnwidth]{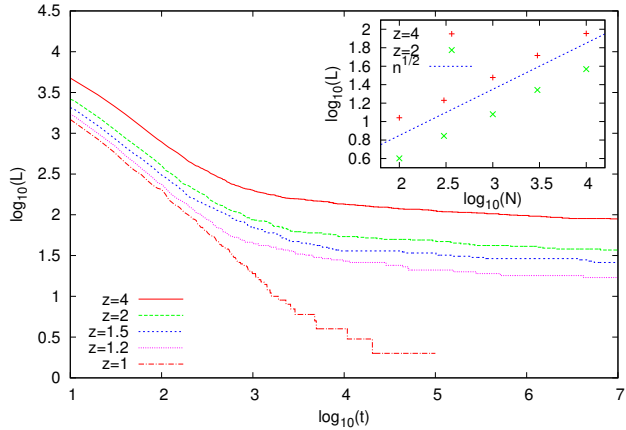}
\caption{Time dependence of the size of the core (log-log scale); calculations were made for $N=10^4$.
Inset shows that for $z=2$ and 4 the 'steady-state' size of the core increases approximately as $N^{1/2}$. To generate the network of fractional $z$ $(1<z<2)$,  with probability $2-z$ we created one out-link and with probability $z-1$ we created two such links for each agent. Thus fractional $z$ has only an average sense.}
\label{time-size-size}
\end{figure}

A more detailed analysis shows, however, that the core size $L$ slowly diminishes in time (Fig.~\ref{time-size-size}). This is not surprising since once an agent looses all of its in-links, it cannot get them back because the probability to be selected in a rewiring process is proportional to the current number of in-links (which is 0 for such an agent). Although the process of diminishing of $L$ is irreversible, it is extremely slow for $z>1$.  Only at $z=1$, this process is considerably faster and in a large time interval consistent with $t^{-1}$ decay. Such a slow decay for $z>1$ suggests that at long (but not infinitely long) time, the core is almost in a  steady state and has a certain size. Numerical calculations show that for $z=2$ and 4 it increases with the system size approximately as $N^{1/2}$ (inset in Fig.~\ref{time-size-size}).   
%%%%%%%%%%%%%%%%%%%%%%%%%%%%%%%%%%%%%%%%%%%%%%%
%%%%%%%%%%%%%%%%%%%%%%%%%%%%%%%

\begin{figure}

\includegraphics[width=\columnwidth]{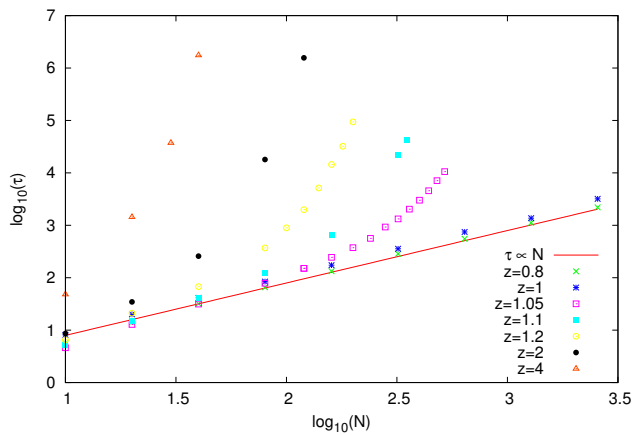}
\caption{The system size dependence of the time $\tau$ to condensate, ie., to reach the core size $z+1$. Let us notice that even for $z$ slightly larger than 1,  $\tau$ shows a rapid increase.}
\label{size-tau}
\end{figure}

Some insight into the stability of the core can be obtained from the analysis of the time~$\tau$ needed for the system to condensate, i.e., for a given $z$ to reach the core size $z+1$ (which is the smallest core size that the system can reach). Numerical calculations show that for $z>1$, $\tau$ exhibits a fast, possibly exponential, increase with the system size (Fig.~\ref{size-tau}). And again,  a slower increase ($\sim N$) is obtained only at $z=1$.  
%%%%%%%%%%%%%%%%%%%%%%%%%%%%%%%%%%%%%%%%%%%%%%%
%%%%%%%%%%%%%%%%%%%%%%%%%%%%%%%

Our results in Figs.~\ref{time-size-size}--\ref{size-tau} show that rewiring for $z=1$ leads to  a rather fast  condensation while for $z>1$ the dynamics  is basically trapped in  a core of size $\sim N^{1/2}$. Even though the condensed state could be reached in principle, for large $N$ and $z>1$ it virtually never happens. The situation is reminiscent of some models with the so-called absorbing states: for some values of control parameters, the absorbing state of the dynamics is basically unavailable and the model remains in the active phase (the lifetime of which in that regime is also exponentially divergent with the system size) \cite{absorbing}.

A network structure with  a nearly stable core suggests an explanation of the generic divergence of susceptibility that we reported in the previous section. Indeed, since agents are influenced only by agents from the core, on average there are $N^{1/2}$ agents that are influenced by a single agent belonging to the core.  Considering core agents as independent (and influencing $N^{1/2}$ other agents), we easily obtain that $\chi \sim N^{1/2}$. Numerical results (Fig.~\ref{susc}) suggest a faster increase at low temperatures (for $T=3$, we obtained $\chi\sim N^{0.91}$) that most likely result from some correlations between core agents. Another factor affecting our simple estimations of the divergence of $\chi$ might be some heterogeneity of the core (Fig.~\ref{conf-core}). 
%%%%%%%%%%%%%%%%%%%%%%%%%%%%%%%%%%%%%%%%%%%%%%%%%%%%%%
\section{conclusions and remarks}
In the present paper we have shown that preferential rewiring changes an Ising ferromagnet, which has a single critical point, into a model with robust critical behaviour. The rewiring mechanism that we used is supposed to mimick the behaviour  of financial agents who try to follow their neighbours but at the same time have also some freedom to choose the ones to follow. We assume that the preference in the rewiring process is proportional to the number of  in-links of a given agent. It is thus not a (more or less) objective measure of its performance but solely how the agent is perceived by the population of other agents. Similar recipes turned out to be successful in, e.g., some page rank algoritms used by search engines \cite{brin} or various recommendation systems \cite{amazon}.

Our model is of some interest from the statistical-mechanics point of view. Due to out-homogeneity, we could write a simple mean-field-like equation (Eq.~(\ref{mfa})) that can be used to obtain the magnetization of the model. We argued, however, that for the present model in the non-rewired version this equation should be exact, and numerical simulations provide a very strong support for the claim.  A very good agreement with this equation was obtained also for the rewired case, but only in the low-temperature regime. We made some attempts to explain why rewiring invalidates this equation at  higher temperatures and at the same time leads to the divergence of suceptibility and criticality. In our opinion this is related with the formation of  a relatively small subset of agents that retain some in-links and  thus drive the entire system.

The change of the dynamical regime in the rewired version at $z=1$ is also of some interest. It is tempting to associate the change with some percolation transition that for random graphs is known to  take place at $z=1$ \cite{random-graphs}. However, rewiring redistributes links in a highly nonrandom fashion and a possible relation with random graphs is by no means obvious.

We  hope that our model might be useful also in the econophysics context. Relatively simple rules that generate a robust criticality might serve as a starting point for further modifications and analysis. 
For example, one might consider a model where an agent that no longer has any in-links still retains some (small) status $\epsilon$ and can be thus selected during the rewiring process. We expect that for small (possibly $N$-dependent) $\epsilon$ such a model would be similar  to our ($\epsilon=0$) model, but  certainly numerical simulations would be needed to support such a claim.
One of the important stylized facts that apparently is missing  in our model is volatility clustering. One might hope that some extensions where agents, for example, try to be in the minority (like in the so-called minority games \cite{minority}) or use more sophisticated strategies (like  'fundamentalist', 'trend follower' or 'noise trader') will provide a more realistic description of financial markets and at the same time will retain simplicity of the model .

Acknowledgements: The research for this work was supported by NCN grant 2013/09/B/ST6/02277 (A.L.), NCN grant 2011/01/B/HS2/01293 (D.L.) and Ministry of Science and  Higher Education grant N N202 488039 (K.G.). 
%%%%%%%%%%%%%%%%%%%%%%%%%%%%%%%%%%%%%%%%%%%%%%%%%%%%

%%%%%%%%%%%%%%%%%%%
%%%%%%%%%%%%%%%%%%%%%%%%%%%%%%%%%%%%%%%%%%%%%%%%%%%%%%%%%%%%%%%%%%%%%%%%%%%%%%%
%%%%%%%%%%%%%%%%%%%%%%%%%%%%%%%%%%%%%%%%%%%%%%%%%%%%%%%%%%%%%%%%%%%%%%%%%%%%%%%
\end {document}